\documentclass[aps,prb,amsmath,amssymb,superscriptaddress,nofootinbib,longbibliography,onecolumn]{revtex4-2}
\usepackage{graphicx}
\usepackage{mathrsfs}
\usepackage{bm}
\usepackage{textcomp,color}
\usepackage{xcolor}
\usepackage[colorlinks=true,urlcolor=blue,citecolor=blue,linkcolor=blue,bookmarks=false, pdfstartview={FitH}]{hyperref}
\usepackage[capitalize]{cleveref} 
\usepackage{amsmath}
\usepackage[normalem]{ulem} 

\newcommand{\beq}{\begin{equation}}
\newcommand{\eeq}{\end{equation}}
\newcommand{\bea}{\begin{eqnarray}}
\newcommand{\eea}{\end{eqnarray}}
\newcommand{\bse}{\begin{subequations}}
\newcommand{\ese}{\end{subequations}}
\newcommand{\nn}{\nonumber}
\newcommand{\bwt}{\begin{widetext}}
\newcommand{\ewt}{\end{widetext}}

\newcommand{\e}{\epsilon}

\newcommand{\bk}{{\bf k}}

\newcommand{\br}{{\bf r}}
\newcommand{\bv}{{\bf v}}

\newcommand{\bE}{{\bf E}}

\newcommand{\bJ}{{\bf J}}

\begin{document}

\title{Effects of electron correlation on resonant Edelstein and inverse-Edelstein effects}
\author{Mojdeh Saleh}
\affiliation{Department of Physics, Concordia University, Montreal, QC H4B 1R6, Canada}
\author{Abhishek Kumar}
\affiliation{Département de Physique and Institut Quantique, Université de Sherbrooke, Sherbrooke, Québec J1K 2R1, Canada}
\affiliation{National High Magnetic Field Laboratory, Tallahassee, Florida 32310, USA}
\affiliation{Department of Physics, Florida State University, Tallahassee, Florida 32306, USA}
\author{Dmitrii L. Maslov}
\affiliation{Department of Physics, University of Florida, Gainesville, Florida, 32611, USA}
\author{Saurabh Maiti}
\affiliation{Department of Physics, Concordia University, Montreal, QC H4B 1R6, Canada}
\affiliation{Centre for Research in Multiscale Modelling, Concordia University, Montreal, QC H4B 1R6, Canada}
\date{\today}
\begin{abstract}
Spin-orbit coupling in systems with broken inversion symmetry gives rise to the Edelstein effect, which is the induced spin polarization in response to an applied electric field or current, and the inverse Edelstein effect, which is the induced electric current in response to an oscillatory magnetic field or spin polarization. At the same time, an interplay between spin-orbit coupling and electron-electron interaction leads to a special type of collective excitations--chiral-spin modes--which are oscillations of spin polarization in the absence of a magnetic field. As a result, both Edelstein and inverse Edelstein effects exhibit resonances at the frequencies of spin-chiral collective modes. Here, we present a detailed study of the effect of electron correlation on the Edelstein and inverse Edelstein effects in a single-valley two-dimensional electron gas and a multi-valley Dirac system with proximity-induced spin-orbit coupling. While the chiral-spin modes involve both in-plane and out-of-plane oscillations of spins, we show that only the in-plane modes are responsible for the above resonances. In the multi-valley system, electron correlation splits the in-plane modes into two. We also study the spectral weight distribution between the two resonances over a large parameter space of intra- and inter-valley interactions.
\end{abstract}

\maketitle
\tableofcontents
\section{Introduction}\label{Sec:Introduction}
In systems with inversion symmetry, electron spin couples to its momentum in two ways. One is Mott skew scattering \cite{Mott1965}, when
electrons scatter from heavy-atom impurities with an amplitude that depends on the angle between the momentum and spin. The other one is the Elliot-Yafet mechanism \cite{Elliot1954,YAFET1963}, when Bloch electrons residing in a spin-orbit coupled energy band scatter from (not necessarily heavy) impurities. On the other hand, broken inversion symmetry lifts the two-fold degeneracy of electron states while still preserving the Kramers degeneracy. Some examples of spin-orbit coupling (SOC) in systems with broken inversion symmetry are Dresselhaus SOC \cite{Dresselhaus1955}, Rashba SOC \cite{Bychkov1984}, and valley-Zeeman (VZ) (also known as Ising) SOC \cite{Kormanyos2014}. The spin-orbit interaction in both scenarios (with and without inversion symmetry breaking) leads to several spin-dependent transport phenomena,  such as the spin Hall effect \cite{Dyakonov1971a,DYAKONOV1971b} (induced spin accumulation transverse to the electric current)
and its inverse \cite{Averkiev1983} (induced current flow from non-uniform spin injection), and current-induced spin polarization in systems with linear-in-momentum spin-splitting \cite{Aronov:1989}, known in the context of two-dimensional electron gases (2DEGs) as the Edelstein effect \cite{EDELSTEIN1990}, and its inverse \cite{shen2014}. Apart from these steady-state effects, there is also the electric-dipole spin resonance (EDSR) effect \cite{Rashba2003,Efros2006,Duckheim2006} (where the spin resonance is driven electrically),
dynamical spin accumulation at the 2DEG boundaries \cite{Duckheim2009}, and the chiral-spin resonance (CSR) which is an electron spin resonance (ESR) at the frequency related to intrinsic spin-orbit splitting
in the absence of the magnetic field \cite{shekhter2005,maiti2016,kumar2021}. 

While the observation of the steady-state effects require some mechanism of either momentum or spin relaxation, the resonant effects require coupling of the electromagnetic field to spin collective modes 
of the system. When the spin splitting is dominated by the external Zeeman field,\footnote{By ``Zeeman field'', we mean a magnetic field that couples to electron spins but not to their orbital motion. It can be an in-plane magnetic field applied to a two-dimensional electron system or the effective exchange field from a ferromagnet in proximity with a non-magnetic electron system.} the relevant collective mode is a Silin-like 
mode \cite{Silin1958} near the Larmor frequency, which can be observed not only via ESR but also via EDSR effect, i.e., a spin-orbit assisted coupling between the electric field of the electromagnetic wave and electron spins \cite{Rashba2003,Efros2006}. The intrinsic spin splitting also leads to collective modes known as chiral-spin modes (CSMs) with frequencies typically in the THz range \cite{shekhter2005,ashrafi2012,maiti2015a}, which 
can be observed via both ESR \cite{maiti2016} and EDSR effects \cite{kumar2021} even in the absence of the Zeeman field.

In Ref.~\cite{MojdehA}, we showed that CSMs also lead to resonances in two more effects namely, in the dynamical Edelstein effect and its inverse. This means that an oscillatory electric field induces not only oscillatory electric current but also oscillatory magnetization. Likewise, an oscillatory Zeeman field induces not only oscillatory magnetization but also an oscillatory electric current. These additional responses were termed as ``cross-responses". In this article, we analyze the effect of electron correlation on cross-responses. We do so by using the Landau kinetic equation for a Fermi liquid that allows one to derive the full tensor of direct and cross-responses. We consider two prototypical systems: a single-valley 2DEG with Rashba or Dresselhaus SOC, such as commonly encountered in semiconductor heterostructures, and a doped multi-valley Dirac system with proximity-induced SOC of both Rashba and valley-Zeeman types, such as in graphene on a transition-metal-dichalcogenide (TMD) substrate \cite{Wang2015,morpurgo:prx}. The primary effect of electron correlations in a single-valley 2DEG is to renormalize the resonance frequencies. In a multi-valley system, however, correlations also lead to splitting of the resonances in both direct \cite{kumar2021} and, as shown in this paper, cross-responses. In particular, we show that for a generic choice of interaction parameters the spectral weights of the two peaks in a split resonance differ significantly. The lower energy mode is shown to carry most of the spectral weight in all responses except the electrical conductivity. Here, there exists a significant parameter regime in the space of intra- and inter-valley correlations where the spectral weights are rather evenly distributed between the two modes.

The rest of the paper is organized as follows. In Sec.~\ref{Sec:KinEq} we discuss the formalism of the Landau kinetic equation on general grounds. The resonant responses of a 2DEG and a multi-valley system are discussed in Secs.~\ref{Sec:2DEG} and \ref{Sec:Dirac}, respectively. The effect of electron correlations on both is discussed in Sec.~\ref{Sec:Interactions}. The spectral weight analysis is presented in Sec. \ref{sec:analysis} and  our conclusions are summarized in Sec.~\ref{Sec:Conclusion}.

\section{The Landau kinetic equation}\label{Sec:KinEq}
Following the method summarized in Ref. \cite{Maslov2022}, we employ the Landau kinetic equation \cite{landau1980} in the following form (here and thereafter we set $\hbar=1$):
\bea\label{Eq:LKE}
\partial_t\hat n+i[\hat \epsilon,\hat n] + \frac 12\{\partial_{\br}\hat n,\partial_{\bk}\hat \epsilon\}- \frac 12\{\partial_{\bk}\hat n,\partial_{\br}\hat \epsilon\}=\hat I_{\rm coll},
\eea
where $\hat \epsilon\equiv \hat  \epsilon(\bk,t)$ is the quasiparticle energy and $\hat n\equiv \hat n(\bk,t)$ is the non-equilibrium distribution function. Both $\hat\epsilon$ and $\hat n$ are matrices of the corresponding dimensionality in the spin and valley subspaces, which are parameterized as
\bse
\bea
\hat \epsilon\equiv \hat  \epsilon(\bk,t)&=&\hat H_0(\bk)+\hat H_{\rm SOC}(\bk) 
+\hat \epsilon_{\rm LF}(\bk,t) 
+ \delta \hat\epsilon_{\rm ext}(t),\label{eq:definitions_eps}\\
\hat n\equiv \hat n(\bk,t)&=&\hat n_0(\bk) +\hat n_{\rm SOC}(\bk)
+\delta \hat n(\bk,t).\label{eq:definitions_n}
\eea
\ese
In Eq.~\eqref{eq:definitions_eps},
$\hat H_0(\bk)$ is 
the single-particle Hamiltonian in the absence of SOC, $\hat H_{\rm SOC}(\bk)$ describes the SOC,
\beq\label{eq:LF}
\hat\epsilon_{\rm LF}(\bk,t)=\int_{\bk'} {\rm Tr} [\hat F(\bk,\bk')\delta \hat n(\bk',t)] \eeq
describes the effect of interaction between quasiparticles via the Landau functional (LF) $\hat F(\bk,\bk')$ with $\int_{\bk}\equiv \int d^2k/(2\pi)^2$, and $\delta\hat \e_{\rm ext}(t)$ is a correction to the energy due to the driving oscillatory electric and Zeeman fields. Explicit forms of $\hat{H}_0(\bk)$ and $\hat{H}_{\text{SOC}}(\bk)$ will be presented later in Secs.~\ref{Sec:2DEG} and \ref{Sec:Dirac}, where specific systems will be discussed. The Landau functional is taken to be the same for a given system in the absence of SOC. This means that SOC is assumed to be weak, such that $\lambda_{\rm SOC}/\mu\ll1$, where $\mu$ is the chemical potential and $\lambda_{\rm SOC}$ is the energy splitting of spins at the Fermi level, and treated perturbatively. Next, in Eq.~\eqref{eq:definitions_n}, $\hat n_0(\bk)$ is the equilibrium electron distribution, $\hat n_{\rm SOC}(\bk)=n'_0\hat H_{\rm SOC}(\bk)$ is the correction due to SOC with $n'_0\equiv \partial_{\epsilon} n_0$, and  $\delta \hat n(\bk,t)$ is the non-equilibrium part of $\hat n$. Note that the deviation from equilibrium enter both directly, via $\delta \hat n(\bk,t)$, and indirectly, via the corresponding change in the quasiparticle energy, as specified by Eq.~\eqref{eq:LF}. Finally, the collision term on the right-hand side of Eq.~\eqref{Eq:LKE} accounts for relaxation processes. 
A microscopic description of relaxation processes is beyond the scope of this paper. At the phenomenological level, $\hat I_{\rm coll}$ can be decomposed into the parts corresponding to relaxation of different degrees of freedom, i.e., charge, in-plane and out-of-plane components of spins, and valley polarization. A discussion of these degrees of freedom will be expanded on in Secs. \ref{Sec:2DEG} and \ref{Sec:Dirac}. For each component, one can introduce a phenomenological relaxation time \cite{RAKITSKII2025}. In the ballistic limit, where $\lambda_{\rm SOC}\tau_{\min}\gg1$ ($\tau_{\min}$ being the shortest relaxation time), the net effect is only to introduce an effective broadening of the resonances. To avoid dealing with the multitude of phenomenological parameters, we set all the relaxation times to be equal to the momentum relaxation time $\tau$ and model the collision integral as $\hat I_{\rm coll}=-\delta\hat n(\bk)/\tau$. Microscopically, such a case is encountered--up to numerical coefficients--in the ballistic limit of the D'yakonov-Perel' spin-relaxation mechanism due to scattering by short-range impurities in the presence of Rashba and/or Dresselhaus SO coupling \cite{DYAKONOV1972}. As a consequence of choosing to work in the ballistic regime, we caution the reader that the limit of $\Omega=0$ cannot be taken in the results presented in this paper.

The driving electric ($E$) and Zeeman ($B$) fields are assumed to be weak and spatially uniform, such that the kinetic equation can be linearized in $\delta\hat n$ and $\delta\hat \e_{\rm ext}(t)$. Also, in accord with an assumption of weak SOC, it will be also  linearized in $\hat H_{\rm SOC}$. In this regard, a term $\propto\hat H_{\rm SOC}\hat \e_{\rm ext}$ is an allowed term as it is still linear in its constituents. If the external driving force is due to the $E$-field, the linearized kinetic equation takes the form\footnote{Here, we ignore the term $\propto n''_0\hat H_{\rm SOC}\mathbf{v}_{\bk}\cdot \mathbf E_0$ as it does not affect the resonant behavior that we wish to model.}: 
\beq\label{eq:EdriveGenE}
\partial_t\delta\hat n + i[\hat H_{\rm SOC},\delta \hat n]-in'_0[\hat H_{\rm SOC},\int_{\bk'}\hat F_{\bk\bk'}\delta\hat n_{\bk'}]=e n'_0(\hat {\mathbf v}_{\bk}+\hat{\mathbf v}_{\rm SOC})\cdot\mathbf E_0 e^{-i\Omega t}+\hat I_{\rm coll},
\eeq
where $\mathbf{E}_0e^{-i\Omega t}$ is the applied oscillatory electric field which has only in-plane components, and $
\hat{\bv}_\bk \equiv \partial\hat{H}_0/\partial\bk$ and $\hat{\bv}_\text{SOC} \equiv \partial\hat{H}_\text{SOC}/\partial\bk$ are the parts of the velocity operator with and without SOC, respectively.
The coupling of the electric field to electron spins, i.e., the EDSR effect, occurs via the $e\hat n_0' \hat{\mathbf v}_{\rm SOC}\cdot\mathbf E_0 e^{-i\Omega t}$ term on the right-hand side of Eq.~\eqref{eq:EdriveGenE}.

For the $B$-driven case, the equation takes the form
\beq\label{eq:EdriveGenB}
\partial_t\delta\hat n + i[\hat H_{\rm SOC},\delta \hat n]-in'_0[\hat H_{\rm SOC},\int_{\bk'}\hat F_{\bk\bk'}\delta\hat n_{\bk'}]=-in'_0\frac{g\mu_{\rm{B}}}{2}[\mathbf B_0\cdot\hat{{\mathbf s}}+B_z\hat s_z,\hat H_{\rm SOC}]e^{-i\Omega t} +\hat I_{\rm coll},
\eeq
where $(\mathbf{B}_0,B_z)e^{-i\Omega t}$ is the applied oscillatory Zeeman field. Throughout the paper, we will use bold symbols to denote in-plane vectors, while the $z$-component will be labeled explicitly. Note that a spatially-uniform but oscillatory Zeeman field does not induce magnetization in the absence of SOC, because the spin susceptibility vanishes in this limit as dictated by spin conservation. Accordingly, the driving term in the right-hand side of Eq.~\eqref{eq:EdriveGenB} vanishes for $\hat H_{\rm SOC}=0$.

We will be interested in two observables: the electric current density which one deduces from the continuity equation, and magnetization. These are given by \cite{nozieres,physkin}
\bse
\bea
{\bf J}&=&-\frac{e}{2}\int_{\bk}{\rm Tr}\left[\{\partial_\bk(\hat H_0+\hat H_{\rm SOC}),(\delta\hat n-\hat n'_0\hat \e_{\rm LF})\}\right],\label{eq:responsesJ}\\
{\bf M}&=&-\frac{g\mu_{\rm{B}}}{2}\int_\bk{\rm Tr}[\hat{\mathbf s}\delta\hat n],~~M_z=-\frac{g\mu_{\rm{B}}}{2}\int_\bk{\rm Tr}[\hat{s}^z\delta\hat n],\label{eq:responsesM}
\eea 
\ese
where $-e$ is the electron charge, $g$ is the effective Land\'e $g$-factor, $\mu_{\rm{B}}$ is the Bohr-magneton, and $\hat s^\alpha$ with $\alpha\in\{x,y,z\}$ are matrices in the spin space. The traces are taken over both the spin and valley subspaces. All the parameters of the single-particle Hamiltonians in the above equations~-$g$, $v_{\rm SOC}$, etc.-~are assumed to include Fermi-liquid renormalizations. To simplify notations, we will use the same symbols for bare and renormalized quantities.

\section{Single-valley two-dimensional electron gas: non-interacting limit}\label{Sec:2DEG}
\subsection{Cross-responses in the presence of  Rashba spin-orbit coupling}\label{subsec:Rashba}
We begin with the Rashba type of SOC in which case
\bea\label{Eq:2deg}
\hat H_0=\left(\dfrac{\bk^2}{2m}-\mu\right)\hat s_0,~~~\hat H_{\rm SOC}=v_{\rm R}(\bk\times\hat{\mathbf s})\cdot{\mathbf z}.
\eea
The `$~\hat{~}~$' now indicates a matrix in the $2\times 2$ spin space and $v_{\rm R}$ is the Rashba coupling constant with units of velocity. Correspondingly,
\bea\label{Eq:2degvel}
\hat{\mathbf v}_{\bk}=\dfrac{{\bk}}{m}\hat s_0,~~~\hat{\mathbf v}_{\rm SOC}=v_{\rm R}~\bk_u\times\hat{\mathbf s},
\eea
where $\bk_u$ is the unit vector along $\bk$. Solving the kinetic equation becomes considerably simpler if the $2\times 2$ distribution function is expanded over the Pauli matrices in the``chiral'' basis 
\cite{shekhter2005,kumar2017}:
\beq\label{eq:LKE_2deg}
\delta\hat n(\bk)=-\delta(\epsilon-\epsilon_F)u_i(\bk)\hat c^i,~~~~i\in\{0,1,2,3\},
\eeq
where ${\hat c}^0={\hat{s}}^0,~~{\hat c}^1=-{\hat{s}}^z,~~{\hat c}^2={\hat{s}}^x\cos\theta+{\hat{s}}^y\sin\theta~$, ${\hat c}^3={\hat{s}}^x\sin\theta-{\hat{s}}^y\cos\theta$, $\theta$ is azimuthal angle of $\bk$, and $-\delta(\epsilon - \epsilon_F)$ is the zero temperature limit of $\partial_\epsilon n_0$. The $i$'s label the degrees of freedom of the system. Due to the $\delta$-function, the functions $u_i(\bk)$ are projected onto the Fermi surface and depend only on $\theta$: $u_i(\bk)=u_i(\theta)$. Since the Rashba SOC is rotationally-symmetric, the kinetic equations for harmonics of $u_i(\theta)$ with different projections of the angular momentum onto the $\hat z$-axis should decouple. It is then convenient to expand the functions $u_i(\theta)$ 
over a complete basis of the eigenstates of the angular momentum operator as $u_i(\theta)=\sum_{m=-\infty}^\infty u_i^{(m)}e^{im\theta}$. In terms of $u^{(m)}_i$, the current density is reformulated as
\bea\label{eq:J2DEG}
J_x&=&e\nu^T_F\left[v_{\rm{F}}\frac{u^{(-1)}_0+u^{(1)}_0}{2}+\frac{\lambda_{\rm R}}{2k_{\rm{F}}}\left(\frac{u^{(-1)}_2-u^{(1)}_2}{2i}-\frac{u^{(-1)}_3+u^{(1)}_3}{2}\right)\right],\nn\\
J_y&=&e\nu^T_F\left[v_{\rm{F}}\frac{u^{(-1)}_0-u^{(1)}_0}{2i}-\frac{\lambda_{\rm R}}{2k_{\rm{F}}}\left(\frac{u^{(-1)}_2+u^{(1)}_2}{2}+\frac{u^{(-1)}_3-u^{(1)}_3}{2i}\right)\right],
\eea 
where $v_{\rm{F}}$ is the Fermi velocity, $k_{\rm{F}}$ is the Fermi momentum, $\lambda_{\rm R}\equiv 2v_{\rm R} k_{\rm{F}}$ is the energy splitting due to SOC at the Fermi level, and $\nu_F^T=m/\pi$ is the total density of states at the Fermi level. The first term in the current density is the usual contribution from free carriers that gives the Drude response, while the second term is due to SOC. Likewise, the magnetization is reformulated as
\bea\label{eq:M}
M_x&=&\frac{g\mu_{\rm{B}}\nu^T_F}2\left(\frac{u_2^{(-1)}+u_2^{(1)}}{2}+\frac{u_3^{(-1)}-u_3^{(1)}}{2i}\right),\nn\\
M_y&=&\frac{g\mu_{\rm{B}}\nu^T_F}2\left(\frac{u_2^{(-1)}-u_2^{(1)}}{2i}-\frac{u_3^{(-1)}+u_3^{(1)}}{2}\right),\nn\\
M_z&=&\frac{g\mu_{\rm{B}}\nu^T_F}2(-u_1^{(0)}).
\eea

Taking the Fourier transform of the kinetic equation with respect to $t$, we find the decoupled equations of motion for $E$-field driving to be:
\bea\label{eq:EoM2DEGEfield}
i\Omega u_0^{(m)}&=&-\frac{ev_{\rm{F}}}{2}\Big[\left( \delta_{m,1} + \delta_{m,-1} \right) E_x - i\left( \delta_{m,1} - \delta_{m,-1} \right) E_y \Big],\nn\\
i\Omega u_1^{(m)}+\lambda_{\rm R}u_2^{(m)}&=&0,\nn\\
-\lambda_{\rm R}u_1^{(m)}+i\Omega u_2^{(m)}&=&ie\dfrac{\lambda_{\rm R}}{4k_{\rm{F}}} \Big[\left( \delta_{m,1} - \delta_{m,-1} \right) E_x-i \left( \delta_{m,1} + \delta_{m,-1} \right) E_y \Big],\nn\\
i\Omega u_3^{(m)}&=&e\dfrac{\lambda_{\rm R}}{4 k_{\rm{F}}} \Big[\left( \delta_{m,1} + \delta_{m,-1} \right) E_x - i\left( \delta_{m,1} - \delta_{m,-1} \right) E_y \Big].
\eea
Here, $E_x$ and $E_y$ are the in-plane components of $\mathbf E_0$. It is evident that the electric field couples only to the $m=\pm1$ harmonics. This makes sense as $E$ is expected to excite dipoles which belong to $|m|=1$ angular momentum channels. It is also evident from Eq.~\eqref{eq:EoM2DEGEfield} that left-circularly polarized light drives one of the $m=\pm1$ channels while the right-circularly polarized light drives the other. We also note that $M_z$ is given in terms of the $m=0$ harmonic of $u_i$ while the in-plane magnetization are given in terms of the $m=\pm1$ harmonics of the same. Since the $E$-field  drives only the $m=\pm 1$ harmonics, we see that it also induces the $M_x$ and $M_y$ components of the magnetization (of amplitude $\propto \lambda_{\rm R}$) but not $M_z$.\footnote{That the electric field induces only the in-plane component of magnetization is valid only if the relaxation mechanism corresponds to isotropic scattering, which is what we have assumed. If scattering is anisotropic, the out-of-plane component is induced as well \cite{Engel:2007}.} Here and henceforth in this article, all occurrences of $i\Omega$ should be seen as $i\Omega-1/\tau$, unless otherwise stated.

For $B$-field driving, our equations read
\bea\label{eq:EoM2DEGBfield}
i\Omega u_0^{(m)}&=&0,\nn\\
i\Omega u_1^{(m)}+\lambda_{\rm R}u_2^{(m)}&=&\dfrac{g\mu_{\rm{B}}\lambda_{\rm R}}{4}\Big[\left( \delta_{m,1} + \delta_{m,-1} \right) B_x - i\left( \delta_{m,1} - \delta_{m,-1} \right) B_y \Big],\nn\\
-\lambda_{\rm R}u_1^{(m)}+i\Omega u_2^{(m)}&=&\dfrac{ g\mu_{\rm{B}}\lambda_{\rm R}}{2} \delta_{m,0} B_z,
\nn\\
i\Omega u_3^{(m)}&=&0.
\eea 

Similar to the $E$-field case, the $m=\pm 1$ harmonics are driven by an in-plane $B$-field, which will translate to an induced electric current. Also like the $E$-field case, the left/right polarized $B$-field couples to one of the $m=\pm 1$ harmonics. However, unlike the $E$-field case, the $B_z$ component does drive the $m=0$ fluctuation which will manifest in the $M_z$ response.

Solving the linear system \eqref{eq:EoM2DEGEfield} we find the  electric current induced by the $E$-field
\beq\label{eq:j2D-e}
\begin{split}
\bJ=  
\underbrace{
\frac{i\sigma_0}{\tau}\left[\frac1{\Omega }+\frac1\Omega\left(\frac{\lambda_{\rm R}}{4\mu}\right)^2R^{\rm 2D}_{\rm JE}(\Omega)\right]
}_{\equiv\sigma(\Omega)}
{\bE}_0, \,\,\,\, {\rm where}\;R_\text{JE}^{\rm 2D}(\Omega) = 2+\frac{\lambda_{\rm R}^2}{\Omega^2 - \lambda_{\rm R}^2},
\end{split}
\eeq
and 
\bea
\sigma_0=e^2\nu_F^Tv_{\rm{F}}^2\tau/2\label{sigma0}
\eea is the \emph{dc} Drude conductivity. The prefactor of ${\bf E}_0$ is identified as the total charge conductivity $\sigma(\Omega)$. 
The first, $1/\Omega $ term in the square brackets gives the frequency dependence of the Drude contribution, while the second term represents renormalization of the Drude weight by Rashba SOC 
(via the $\Omega=0$ limit of $R_\text{JE}^\text{2D}$),
as well as the EDSR peak at $\Omega=\lambda_{\rm R}$. Additionally, we see from Eqs. (\ref{eq:J2DEG}) and  (\ref{eq:M}) that the same oscillatory harmonics are present in both $\bf J$ and $\bf M$. Thus, driving with $E$-field produces not only the direct response, i.e., the charge current, but also a cross-response, i.e., magnetization. A straightforward calculation leads to
\bea\label{eq:m2DEGe}
\bf M&=&\underbrace{-\dfrac{i\sigma^{\rm ME}_0}{\Omega\tau}R^{\rm 2D}_{\rm ME}(\Omega)}_{\sigma^{\rm ME}(\Omega)}{\bf E}_0\times \hat{z},\;
M_z=0,\eea
where
\bea
\sigma^{\rm ME}_0 \equiv \frac{g\mu_{\rm{B}}}2\frac{e\lambda_{\rm R}\nu_F^T\tau}{4k_{\rm{F}}}\label{sigma0ME}
\eea
is the steady-state Edelstein conductivity \cite{EDELSTEIN1990}, and $R^{\rm 2D}_{\rm ME}(\Omega)=R^{\rm 2D}_{\rm JE}(\Omega)$ [cf. Eq. \eqref{eq:j2D-e}]. In this case the prefactor of ${\bf E}_0$ is identified with the dynamic Edelstein conductivity $\sigma^{\rm ME}(\Omega)$, which has a resonance at the same frequency $\Omega=\lambda_{\rm R}$ as the charge conductivity. Note that the induced magnetization is transverse to the applied electric field but in the plane of the 2DEG, and that the prefactor of its resonant part is independent of the relaxation time $\tau$.

Similarly, driving with the $B$-field induces not only the direct response, i.e., magnetization, but also cross-response, i.e., charge current.  Explicitly, these responses are given by
\bea\label{eq:mag2DEGb}
\bf M&=&\underbrace{-\frac{\chi_0}{2}R^{\rm 2D}_{\rm MB}(\Omega)}_{\chi_{\parallel}(\Omega)}{\bf B}_0,\;M_z=\underbrace{-\chi_0 R^{\rm 2D}_{\rm MB}(\Omega)}_{\chi_{\perp}(\Omega)}B_z \,\,\,\, R_\text{MB}^{\rm 2D}(\Omega) = R_\text{JB}^{\rm 2D}(\Omega), \nn\\
\bf{J}&=&\underbrace{-\chi_0^{\rm JB} R^{\rm 2D}_{\rm JB}(\Omega)}_{\chi^{\rm JB}(\Omega)}{\bf B}_0\times\hat{z}, \,\,\,\, R_\text{JB}^{\rm 2D}(\Omega) = \frac{\lambda_{\rm R}^2}{\Omega^2-\lambda_{\rm R}^2}
\eea
where
\bea
\chi_0\equiv \frac14 g^2\mu_{\rm{B}}^2\nu^T_F\label{chi0}
\eea
is the static spin susceptibility of a 2DEG in the absence of SOC and 
\bea
\chi_0^{\rm JB}\equiv \frac{\sigma^{\rm ME}_0}{\tau}
=\frac{g\mu_{\rm{B}}}{2}\frac{e\lambda_{\rm R}\nu_F^T}{4k_{\rm{F}}}\label{chi0JB}.
\eea
While the spin susceptibility is isotropic in the absence of SOC, the \emph{dynamic} susceptibility in the presence of SOC is anisotropic, such that its in-plane  component ($\chi_\parallel$) is twice smaller than the out-of-plane  one ($\chi_\perp$). \footnote{The dynamic spin  susceptibility corresponds to the limit of $v_{\rm F}q\ll\Omega$. In the opposite limit of $\Omega\ll v_{\rm F}q\to 0$, the susceptibility is still isotropic even in the presence of SOC, provided that both Rashba subbands are occupied \cite{zak:2010}.} The $B$-field driven charge current  is the resonant inverse Edelstein effect. Note that the current is induced only by the in-plane Zeeman field and in direction perpendicular to ${\bf B}_0$, and that, similar to the case of Edelstein effect, the prefactor of its resonant part is independent of $\tau$. 

\subsection{Cross-responses in the presence of Dresselhaus spin-orbit coupling}\label{subsec:Dresselhaus}
The existence  of cross-responses is not unique to Rashba SOC. Any perturbation that mixes spin and momentum will lead to similar effects but with a different tensor structure of the response functions. As another example, we consider Dresselhaus type of SOC that exists on surfaces of crystals without a bulk inversion  center \cite{Dresselhaus1955,Dyakonov:book}. For a 2DEG on the (001) surface plane, the SO part of the Hamiltonian reads
\bea\label{Eq:2deg-Dresselhaus}
\hat H_{\rm SOC}=v_{\rm D}(\hat s_x k_x-\hat s_yk_y),
\eea
with $v_{\rm D}$ being the Dresselhaus coupling constant. Correspondingly, the SO part of the velocity becomes
\bea\label{Eq:2degvel-D}
\hat{\bf v}_{\rm SOC}=v_{\rm D}(k_{ux}\hat s_x\hat{\bf x}-k_{uy}\hat s_y\hat{\bf y}).
\eea
One can now proceed along the same lines as for the Rashba case. For electric-field driving we find:
\bea\label{eq:jm2dE-D}
\bf J&=&\frac{i\sigma_0}{\tau}\left[\frac1{\Omega }+\frac1\Omega\left(\frac{\lambda_{\rm D}}{4\mu}\right)^2D_{\rm JE}^{\rm 2D}(\Omega)\right]{\bf E}_0,\nn\\
M_x&=&\dfrac{i\sigma^{\rm ME}_0}{\Omega\tau}D_{\rm ME}^{\rm 2D}(\Omega)E_x,\;
M_y=-\dfrac{i\sigma^{\rm ME}_0}{\Omega\tau}D_{\rm ME}^{\rm 2D}(\Omega)E_y.\eea
where $D_{\rm JE}^{\rm 2D}(\Omega)=D_{\rm ME}^{\rm 2D}(\Omega)$ is same as $R_{\rm JE}^{\rm 2D}(\Omega)=R_{\rm ME}^{\rm 2D}(\Omega)$ in Eq. \eqref{eq:j2D-e}, but with $v_{\rm R}\rightarrow v_{\rm D}$, which results in $\lambda_{\rm R}\rightarrow\lambda_{\rm D}$. The same applies to the definition of $\sigma^{\rm ME}_0$ in Eq.~\eqref{sigma0ME}. For Zeeman-field driving, we get
\bea\label{eq:jm2DB-D}
J_x&=&-\frac{\sigma_0^{\rm ME}}{\tau}D_{\rm JB}^{\rm 2D}(\Omega)B_x,\;
J_y=\underbrace{\frac{\sigma_0^{\rm ME}}{\tau}D_{\rm JB}^{\rm 2D}(\Omega)}_{\chi_{\rm JB}}B_y,\nn\\
\bf M&=&\underbrace{-\frac{\chi_0}2D_{\rm MB}^{\rm 2D}(\Omega)}_{\chi_\parallel}{\bf B}_0,\;
M_z=\underbrace{-\chi_0D_{\rm MB}^{\rm 2D}(\Omega)}_{\chi_\perp}B_z,
\eea
where $\chi_0$ is same as in Eq.~\eqref{chi0} and, $D_{\rm JB}^{\rm 2D}(\Omega)
=D_{\rm MB}^{\rm 2D}(\Omega)$ are also same as $R_{\rm JB}^{\rm 2D}(\Omega)=R_{\rm MB}^{\rm 2D}(\Omega)$ [Eq.~\eqref{eq:mag2DEGb}], but, again, with $v_{\rm R}\rightarrow v_{\rm D}$.

In summary, the cross-response tensors for pure Rashba and pure Dresselhaus cases are:
\bea\label{eq:TensorStructures}
\sigma^{\rm ME}_{ab}(\Omega)&=&\begin{cases}
    \frac{-i\sigma_0^{\rm ME}}{\Omega\tau}R^{\rm 2D}_{\rm ME}(\Omega)\begin{pmatrix}
        0&1\\-1&0
    \end{pmatrix},~\text{for Rashba SOC},\\
\frac{-i\sigma_0^{\rm ME}}{\Omega\tau}D^{\rm 2D}_{\rm ME}(\Omega)\begin{pmatrix}
        -1&0\\0&1
    \end{pmatrix},~\text{for Dresselhaus SOC}.
\end{cases}\nn\\
\chi^{\rm JB}_{ab}(\Omega)&=&\begin{cases}
    -\chi_0^{\rm JB}R^{\rm 2D}_{\rm JB}(\Omega)\begin{pmatrix}
        0&1\\-1&0
    \end{pmatrix},~\text{for Rashba SOC},\\
-\chi_0^{\rm JB}D^{\rm 2D}_{\rm JB}(\Omega)\begin{pmatrix}
        1&0\\0&-1
    \end{pmatrix},~\text{for Dresselhaus SOC}.
\end{cases}
\eea
A characteristic difference in the tensor structure of the cross-response functions in the two cases allows for an unambiguous identification of  the dominant type of SOC in a given material. This is to be contrasted with the tensor structure of direct response functions, i.e., conductivity and spin susceptibility, which are longitudinal for both Rashba and Dresselhaus cases. The difference in the tensor structure of cross-response functions could be exploited in transport measurements to extract different components of SOC. 

\section{Multi-valley Dirac system: non-interacting limit}\label{Sec:Dirac}
As a specific realization of the multi-valley system, we consider gated monolayer graphene with two valleys at time-reversal symmetric points of the Brillouin zone. Although intrinsic SOC in graphene is very weak, a strong SOC can be induced by placing graphene on a heavy-element substrate, e.g. TMD. It is known that SOC in this case is of two types: Rashba and valley-Zeeman (VZ), the latter acting as an out-of-plane Zeeman field of polarity alternating between the $K$ and $K'$ valleys \cite{morpurgo:prx}. For energies much smaller than the chemical potential, one can project out the band that does not contain the chemical potential, which leads to the following $4\times 4$ Hamiltonian
\bea\label{eq:GrHam}
\hat H_0=v_{\rm{F}} k \hat \tau^0\hat s^0,~~~\hat H_{\rm SOC}=\frac{\lambda_{\rm R}}{2}\hat \tau^0(\bk_u \times \hat{\bf s})\cdot {\bf z}+\frac{\lambda_{\rm Z}}2\hat \tau^3\hat s^3.
\eea
Here, in addition to the $2\times 2$ spin matrices $\hat s$, there are also $2\times 2$ $\hat{\tau}$ matrices that act in the valley space, and $\lambda_{\rm R}$ and $\lambda_{\rm Z}$ are the Rashba and VZ spin-orbit couplings, respectively, with units of energy. Since a gated system has a finite $k_{\rm{F}}$, we can introduce a velocity $v_{\rm R}\equiv \lambda_{\rm R}/2k_{\rm{F}}$. When comparing the Hamiltonian \eqref{eq:GrHam} with Eq.~\eqref{Eq:2deg}, it is worth noting that the SOC term depends only on the direction of $\bk$ but not on its magnitude, which is the result of projecting out fully occupied bands in the Dirac system. Consequently, the structure of spin-orbit part of the velocity operator
\bea\label{Eq:Gvel}
\hat{\bf v}_{\bk}=v_{\rm{F}}\bk_u\hat\tau^0\hat s_0,~~~\hat{\bf v}_{\rm SOC}=v_{\rm R} (\bk_u\cdot\hat{\bf s}) (\bk_u\times\hat z)\hat\tau^0,
\eea
differs from that in Eq.~\eqref{Eq:2degvel}. Next, the non-equilibrium part of the distribution function is modeled as \cite{Raines2021,kumar2021}
\bea\label{eq:DFGr}
\delta\hat n=-\delta(\epsilon-\epsilon_F)\left[u_0\hat\tau^0\hat c^0+u_i\hat\tau^0\hat c^i+W_i\hat\tau^i\hat c^0+M_{ij}\hat\tau^i\hat c^j\right],~~~~i,j\in\{1,2,3\},
\eea
where $u_i$'s describe the spin-chiral sector in the same way as for 2DEG [cf.~Eq.~\eqref{eq:LKE_2deg}], $W_i$' describe the valley sector, and $M_{ij}$ describe coupled fluctuations in the spin-valley sector. We then perform the harmonic decomposition of these degrees of freedom as
$$u_i(\theta)=\sum_m u^{(m)}_i e^{im\theta},~W_i(\theta)=\sum_m W_i^{(m)} e^{im\theta},~M_{ij}(\theta)=\sum_m M_{ij}^{(m)} e^{im\theta}.$$ 
The electric current and magnetization are still obtained from Eqs.~(\ref{eq:responsesJ}) and (\ref{eq:responsesM}), respectively. For the current, we obtain
\bea\label{eq:JGr}
J_x&=&e\nu^T_F\left[v_{\rm{F}}\frac{u_0^{(-1)}+u_0^{(1)}}{2}+\frac{\lambda_{\rm R}}{2k_{\rm{F}}}\left\{\frac{u^{(-1)}_2-u^{(1)}_2}{2i}
\right\}\right],\nn\\
J_y&=&e\nu^T_F\left[v_{\rm{F}}\frac{u_0^{(-1)}-u_0^{(1)}}{2i}-\frac{\lambda_{\rm R}}{2k_{\rm{F}}}\left\{\frac{u^{(-1)}_2+u^{(1)}_2}{2}\right\}\right],
\eea
where $\nu_F^T \equiv 2k_{\rm{F}}/\pi v_{\rm{F}}$ is the total density of states at the Fermi level.  The expressions for the components of magnetization are same as in the single-valley case [Eq. (\ref{eq:M})] with $\nu_F^T$ corresponding to the multi-valley case now. 
Comparing Eq.~\eqref{eq:JGr} with Eq.~(\ref{eq:J2DEG}), we see that while the Drude part of the current ($\propto v_{\rm{F}}$) remains the same, the SOC induced terms are modified due to a different form of the SOC part of the velocity operator [cf. Eqs.~(\ref{Eq:2degvel}) and (\ref{Eq:Gvel})]. 

The equations of motion for $E$-field driving are:
\bea\label{eq:EomGr1}
i\Omega u_0^{(m)}&=&-\frac{ev_{\rm{F}}}{2}\Big[(\delta_{m,1} + \delta_{m,-1}) E_x - i (\delta_{m,1} - \delta_{m,-1}) E_y \Big],\nn\\
i\Omega u_1^{(m)}+\lambda_{\rm R}u_2^{(m)}&=&0,\nn\\
-\lambda_{\rm R} u_1^{(m)}+i\Omega u_2^{(m)}+\lambda_{\rm Z}M_{3,3}^{(m)}&=&\frac{ie \lambda_{\rm R}}{4k_{\rm{F}}}\Big[(\delta_{m,1} - \delta_{m,-1}) E_x - i (\delta_{m,1} + \delta_{m,-1}) E_y \Big],\nn\\
i\Omega u_3^{(m)}-\lambda_{\rm Z}M_{3,2}^{(m)}&=&0,\nn\\
i\Omega M_{3,1}^{(m)}+\lambda_{\rm R}M_{3,2}^{(m)}&=&0,\nn\\
-\lambda_{\rm R}M_{3,1}^{(m)}+i\Omega M_{3,2}^{(m)}+\lambda_{\rm Z}u_3^{(m)}&=&0,\nn\\
i\Omega M_{3,3}^{(m)}-\lambda_{\rm Z}u_2^{(m)}&=&0.
\eea
When compared to the single valley case, even in the absence of VZ SOC, we see that there are differences in how the degrees of freedom couple to the electric field, e.g., the $u^{(m)}_3$ component does not couple to the electric field. This is a consequence of the modification of the velocity operator [Eq. \eqref{Eq:Gvel}]. Despite this, as we will see soon, the current and magnetization will not be sensitive to this change. Further, with no VZ SOC, the spin-valley sector is also decoupled from the electric field. The presence of VZ SOC, however, couples the spin-chiral and spin-valley degrees of freedom. This suggests that in the presence of VZ SOC, one can use electric field to access fluctuations in the spin-valley sector.

For Zeeman-field driving, the equations of motion read
\bea\label{eq:EomGr2}
i\Omega  u_0^{(m)}&=&0,\nn\\
i\Omega u_1^{(m)}+\lambda_{\rm R}u_2^{(m)}&=&\dfrac{g\mu_{\rm{B}}\lambda_{\rm R}}{4} \Big[(\delta_{m,1} + \delta_{m,-1}) B_x - i (\delta_{m,1} - \delta_{m,-1}) B_y \Big],\nn\\
-\lambda_{\rm R} u_1^{(m)}+i\Omega u_2^{(m)}+\lambda_{\rm Z}M_{3,3}^{(m)}&=&\dfrac{g\mu_{\rm{B}}\lambda_{\rm R}}{2} \delta_{m,0} B_z,\nn\\
i\Omega u_3^{(m)}-\lambda_{\rm Z}M_{3,2}^{(m)}&=&0,\nn\\
i\Omega M_{3,1}^{(m)}+\lambda_{\rm R}M_{3,2}^{(m)}&=&0,\nn\\
-\lambda_{\rm R}M_{3,1}^{(m)}+i\Omega M_{3,2}^{(m)}+\lambda_{\rm Z}u_3^{(m)}&=&-\frac{ig\mu_{\rm{B}}\lambda_{\rm Z}}{4} \Big[(\delta_{m,1} - \delta_{m,-1})B_x - i (\delta_{m,1} + \delta_{m,-1}) B_y \Big],\nn\\
i\Omega M_{3,3}^{(m)}-\lambda_{\rm Z}u_2^{(m)}&=&-\frac{g\mu_{\rm{B}}\lambda_{\rm Z}}{4} \Big[(\delta_{m,1} + \delta_{m,-1}) B_x - i (\delta_{m,1} - \delta_{m,-1}) B_y \Big].
\eea
Without VZ SOC, the spin-chiral sector couples to the Zeeman field in the same way as in the single-valley case, while the spin-valley sector remains decoupled. However, VZ SOC again couples these degrees of freedom. Thus, VZ SOC plays an important role in allowing both the $E$ and $B$ fields to couple to the spin-valley fluctuations.

Solving for the $u^{(m)}_i$ and $M^{(m)}_{ij}$ components and substituting the results into the definitions of the electric current and magnetization, Eqs.~\eqref{eq:responsesJ} and \eqref{eq:responsesM}, we find for the case of electric-field driving:
\bea\label{eq:jGre1}
\bJ&=&\underbrace{\frac{i\sigma_0}{\tau}\left[\frac1{\Omega }+\frac1{\Omega}\left(\frac{\Omega}{2
\mu}\right)^2R^{\rm Di}_{\rm JE}(\Omega)\right]}_{\sigma(\Omega)}{\bf E}_0, \,\,\,\, R_\text{JE}^{\rm Di}(\Omega) = \frac{\lambda_{\rm R}^2}{\Omega^2-\lambda_{\rm R}^2-\lambda_{\rm Z}^2},\nn\\
{\bf M}&=&\underbrace{-\dfrac{i\sigma^{\rm ME}_0}{\Omega\tau}\left(\frac{\Omega}{\lambda_{\rm R}}\right)^2R^{\rm Di}_{\rm ME}(\Omega)}_{\sigma^{\rm ME}(\Omega)} {\bf E}_0\times\hat z,\;
M_z=0,\;R^{\rm Di}_{\rm ME}(\Omega)=R^{\rm Di}_{\rm JE}(\Omega),
\eea
where superscript Di stands for ``Dirac", $\sigma_0$ and $\sigma^{\rm ME}_0$  are the same as in Eqs.~\eqref{sigma0} and \eqref{sigma0ME}, with $\nu_F^T$ re-defined appropriately. 

For Zeeman-field driving, we obtain
\bea\label{eq:BDrive}
\bf M&=&\underbrace{-\frac{\chi_0}2R^{\parallel \rm Di}_{\rm MB}(\Omega)}_{\chi_{\parallel}(\Omega)} {\bf B}_0, \,\,\,\,\,\, 
R^{\parallel \rm Di}_{\rm MB}(\Omega)=\frac{\lambda_{\rm R}^2+2\lambda_{\rm Z}^2}{\Omega^2-\lambda_{\rm R}^2-\lambda_{\rm Z}^2}, \nn\\
M_z&=&\underbrace{-\chi_0R^{\perp \rm Di}_{\rm MB}(\Omega)}_{\chi_{\perp}(\Omega)}B_z, \,\,\,\,\,\, R^{\perp \rm Di}_{\rm MB}(\Omega)= \frac{\lambda_{\rm R}^2}{\Omega^2-\lambda_{\rm R}^2-\lambda_{\rm Z}^2}, \nn\\
\bf J&=&-\chi_0^{\rm JB}R^{\rm Di}_{\rm JB}~{\bf B}_0\times\hat z, \,\,\,\,\,\, R^{\rm Di}_{\rm JB}(\Omega)=\frac{\lambda_{\rm R}^2+\lambda_{\rm Z}^2}{\Omega^2-\lambda_{\rm R}^2-\lambda_{\rm Z}^2}.
\eea
where $\chi_0$ and $\chi_0^{\rm JB}$ are the same as in Eqs.~\eqref{chi0} and \eqref{chi0JB}, respectively. Thus, the cross-responses persist in multi-valley systems. 
In fact, comparing the results in Eqs.~\eqref{eq:jGre1}-\eqref{eq:BDrive} with that in Eqs.~\eqref{eq:j2D-e}-\eqref{eq:mag2DEGb}, we observe that despite the subtle differences in the equations of motion for the fluctuating components, the cross-responses in the non-interacting limit are similar (if not the same) but with an interesting change: the resonance shifts from $\lambda_{\rm R}\rightarrow\sqrt{\lambda_{\rm R}^2+\lambda_{\rm Z}^2}$ (which is expected as that is the new spin splitting in the system). This means that although the resonance factor [$R_{AA'}(\Omega)$] in the cross-response survives when $\lambda_{\rm R}=0$ but $\lambda_{\rm Z}$ is finite, the cross-response itself vanishes when $\lambda_{\rm R}\rightarrow0$, indicating the relevance of Rashba type SOC to finite cross-response.

\section{Dynamic response in the presence of electron-electron interaction\label{Sec:Interactions}}
Electron-electron interaction (\emph{eei}) affects the dynamic response of a system with SOC in a number of ways. The most important one is that it endows electron spins with rigidity with respect to finite-$q$ perturbations. As a result, the various resonances both in the direct and cross sectors become dispersive modes of a FL with SOC \cite{ashrafi2012,Perez2016,maiti2017}. At $q=0$ (which is the subject of this paper), \emph{eei} renormalizes the resonance frequencies. For weak SOC, such renormalization can be expressed in terms of the Landau parameters of the underlying Fermi liquid in the absence of SOC \cite{shekhter2005,kumar2017,kumar2021}. In addition, \emph{eei} leads to damping of the resonances even in the absence of disorder \cite{maiti2015b}. Finally, \emph{eei} is responsible for splitting of both ESR and EDSR peaks in a two-valley Dirac system \cite{kumar2021}.

While the effects of \emph{eei} on direct responses (ESR and EDSR) have been studied extensively in the past (see review \cite{Maslov2022} and references therein), here we extend the analysis to cross-responses. To do so, we come back to Eqs.~(\ref{Eq:LKE}--\ref{eq:LF}) and restore the Landau interaction functional, $\hat\e_{\rm LF}$. Since the Landau functionals are different for a single- and multi-valley cases, we will treat the two separately, in Secs.~\ref{subsec:2D interaction} and \ref{subsec:Dirac interaction}, respectively. Furthermore, \emph{eei} also renormalizes the parameters of a free Hamiltonian, i.e., $v_{F} \to v_{F}^*$, $v_{\rm SOC} \to v_{\rm SOC}^*$ and $g \to g^*$. However, as stated earlier, we shall suppress the asterisks for brevity with the understanding that all parameters in what follows are already renormalized.

\subsection{Single-valley system}\label{subsec:2D interaction}
We start with the standard form of the Landau function for a single-valley system in the absence of both SOC and Zeeman field \cite{landau1980}
\bea\label{eq:ALF}
	\nu_F^T F_{\varsigma_1,\varsigma_2;\varsigma_3,\varsigma_4}(\theta,\theta')=F^S(\theta,\theta')\delta_{\varsigma_1\varsigma_3}\delta_{\varsigma_2\varsigma_4}+F^A(\theta,\theta')\vec{s}_{\varsigma_1\varsigma_3}\cdot\vec{s}_{\varsigma_2\varsigma_4},
\eea
where $F^{S/A}$ are the symmetric and antisymmetric parts describing the interaction in the charge and spin sectors, respectively, and $\vec{A}$ denotes a vector $A$ with three Cartesian components. The angles $\theta,\theta'$ are the azimuthal angles of momenta $\bk,\bk'$ in Eq.~\eqref{eq:LF} with magnitude $k_{\rm F}$;
in a rotationally-invariant system, the Landau function depends only on $\theta-\theta'$. The components of the Landau function can be decomposed into angular harmonics according to $X(\theta,\theta')=\sum_m X_me^{i m(\theta-\theta')}$, where $X\in\{F^S,F^A\}$ and $m$ is the $z$-component  of the angular momentum. It will also be convenient to introduce the following combinations of the Landau parameters:
\bea\label{eq:fs}
&f^{S,(m)}=1+F^S_m,~~~f^{S,(m)}_{+}=1+\frac{F^S_{m-1}+F^S_{m+1}}{2},~~~f^{S,(m)}_{-}=\frac{F^S_{m-1}-F^S_{m+1}}{2};&\nonumber\\
&f^{(m)}=1+F^A_m,~~~f^{(m)}_{+}=1+\frac{F^A_{m-1}+F^A_{m+1}}{2},~~~f^{(m)}_{-}=\frac{F^A_{m-1}-F^A_{m+1}}{2}.&
\eea
In the presence of \emph{eei}, one needs to use the full form of the electric current with the Landau functional taken into account, as given by Eq.~(\ref{eq:responsesJ}). The (temporal Fourier transform of) current density, modified according to Eq.~\eqref{eq:responsesJ}, is then evaluated as
\bea\label{eq:j-int-2d}
J_x&=&e\nu^T_F\left[v_{\rm{F}}f^{S,(1)}\frac{u^{(-1)}_0+u^{(1)}_0}{2}+\frac{\lambda_{\rm R}}{2k_{\rm{F}}}f^{(0)} \left( \frac{u^{(-1)}_2-u^{(1)}_2}{2i}-\frac{u^{(-1)}_3+u^{(1)}_3}{2} \right) \right],\nn\\
J_y&=&e\nu^T_F\left[v_{\rm{F}}f^{S,(1)}\frac{u^{(-1)}_0-u^{(1)}_0}{2i}-\frac{\lambda_{\rm R}}{2k_{\rm{F}}}f^{(0)} \left( \frac{u^{(-1)}_2+u^{(1)}_2}{2}+\frac{u^{(-1)}_3-u^{(1)}_3}{2i} \right) \right].
\eea
The definition of magnetization in Eq.~(\ref{eq:responsesM}) does not involve the Landau function, hence its explicit form remains the same as in the non-interacting case, Eq.~\eqref{eq:M} (modulo ``silent" renormalization of the Hamiltonian parameters).

Next, we follow the same steps as for the non-interacting limit to derive the equations of motion in the presence of \emph{eei}. For electric-field driving those read
\bea\label{eq:EoM2DEG-intE}
i\Omega  u_0^{(m)}&=&-\frac{ev_{\rm{F}}}{2} \Big[(\delta_{m,1} + \delta_{m,-1}) E_x - i (\delta_{m,1} - \delta_{m,-1}) E_y \Big],\nn\\
i\Omega u_1^{(m)}+\lambda_{\rm R}u_2^{(m)}f^{(m)}_{+}+i\lambda_{\rm R}u_3^{(m)}f^{(m)}_{-}&=&0,\nn\\
i\Omega u_2^{(m)}-\lambda_{\rm R}u_1^{(m)}f^{(m)}&=&ie\dfrac{\lambda_{\rm R}}{4k_{\rm{F}}} \Big[(\delta_{m,1} - \delta_{m,-1}) E_x - i (\delta_{m,1} + \delta_{m,-1}) E_y \Big],\nn\\
i\Omega u_3^{(m)}&=&e\dfrac{\lambda_{\rm R}}{4 k_{\rm{F}}} \Big[(\delta_{m,1} + \delta_{m,-1}) E_x - i (\delta_{m,1} - \delta_{m,-1}) E_y \Big],
\eea
while for Zeeman-field driving we obtain
\bea\label{eq:EoM2DEG-intB}
i\Omega  u_0^{(m)}&=&0,\nn\\
i\Omega u_1^{(m)}+\lambda_{\rm R}u_2^{(m)}f^{(m)}_{+}+i\lambda_{\rm R}u_3^{(m)}f^{(m)}_{-}&=&\dfrac{g\mu_{\rm{B}}\lambda_{\rm R}}{4} \Big[(\delta_{m,1} + \delta_{m,-1}) B_x - i (\delta_{m,1} - \delta_{m,-1}) B_y \Big],\nn\\
i\Omega u_2^{(m)}-\lambda_{\rm R}u_1^{(m)}f^{(m)}&=&\dfrac{ g\mu_{\rm{B}}\lambda_{\rm R}}{2} \delta_{m,0} B_z,\nn\\
i\Omega u_3^{(m)}&=&0.
\eea
Compared to the non-interacting case, \emph{eei} renormalizes the already existing couplings and induces some new couplings between the components of the distribution function via the combinations of Landau parameters $f^{(m)}_{\pm}$ and $f^{(m)}$, defined in Eq.~\eqref{eq:fs}, but the driving terms remain the same (up to renormalizations of $v_{\rm{F}},v_{\rm R}, g$, etc.). This will result in the \textit{eei} renormalization and splitting of the resonance frequencies. Solving for the components of $u$, we obtain the following responses for electric-field driving
\bea\label{eq:2DEG-int}
\bf J&=&\underbrace{\frac{i\sigma_0}{\tau}\left[\frac{f^{S,(1)}}{\Omega }+\frac{f^{(0)}}{\Omega}\left(\frac{\lambda_{\rm R}}{4\mu}\right)^2R^{\rm 2D}_{\rm JE}(\Omega)\right]}_{\sigma(\Omega)}{\bf E}_0,\;R^{\rm 2D}_{\rm JE}(\Omega)=2+\dfrac{(f_+^{(1)}+f_-^{(1)})f^{(1)}\lambda_{\rm R}^2 }{\Omega^2-f_+^{(1)}f^{(1)}\lambda_{\rm R}^2},\nn\\
\bf M&=&\underbrace{\dfrac{-i\sigma^{\rm ME}_0}{\Omega\tau}R^{\rm 2D}_{\rm ME}(\Omega)}_{\sigma^{\rm ME}(\Omega)}({\bf E}_0\times\hat{\bf z} ),\;M_z=0,\;R^{\rm 2D}_{\rm ME}(\Omega)=R^{\rm 2D}_{\rm JE}(\Omega),
\eea
and for Zeeman-field driving 
\bea\label{eq:2DEGbc}
\bf{J}&=&\underbrace{-\chi_0^{\rm JB}R^{\rm 2D}_{\rm JB}(\Omega)}_{\chi^{\rm JB}(\Omega)}{\bf B}_0\times\hat{z},~~R^{\rm 2D}_{\rm JB}(\Omega)=\dfrac{f^{(0)}f^{(1)}\lambda_{\rm R}^2}{\Omega^2-f_+^{(1)}f^{(1)}\lambda^2_R},\nn\\
\bf M&=&\underbrace{-\frac{\chi_0}{2}R^{\parallel 2D}_{\rm MB}(\Omega)}_{\chi^\parallel(\Omega)}\mathbf B,~~R^{\parallel 2D}_{\rm MB}(\Omega)=\dfrac{f^{(1)}\lambda_{\rm R}^2}{\Omega^2-f_+^{(1)}f^{(1)}\lambda^2_R},\nn\\
M_z&=&\underbrace{-\chi_0R^{\perp 2D}_{\rm MB}(\Omega)}_{\chi^\perp(\Omega)}B_z,~~R^{\perp 2D}_{\rm MB}(\Omega)=\dfrac{f_+^{(0)}\lambda_{\rm R}^2}{\Omega^2-f^{(0)}f_+^{(0)}\lambda^2_R}.
\eea
As stated, the resonance frequencies are renormalized by \emph{eei}, but differently so for the in- and out-of-plane components of $M$ (for the case of Zeeman-field driving). While both these resonant frequencies show up in the direct response (as was also shown in Refs.  \cite{shekhter2005,kumar2017}), only the frequencies corresponding to the resonance in the in-plane components of $M$ show up in the cross-response.

\subsection{Two-valley system}\label{subsec:Dirac interaction}
The Landau function for a two-valley system can be written as \cite{Raines2021,kumar2021} 
\bea\label{eq:LFGr}
\nu_F^T F_{(\upsilon_1\upsilon_2;\upsilon_3\upsilon_4);(\varsigma_1\varsigma_2;\varsigma_3\varsigma_4)}(\theta,\theta')&=&F^S(\theta,\theta')\delta_{\upsilon_1\upsilon_3}\delta_{\upsilon_2\upsilon_4}\delta_{\varsigma_1\varsigma_3}\delta_{\varsigma_2\varsigma_4}\nn\\
&&+F^A(\theta,\theta')\delta_{\upsilon_1\upsilon_3}\delta_{\upsilon_2\upsilon_4}(\vec{s}_{\varsigma_1\varsigma_3}\cdot\vec{s}_{\varsigma_2\varsigma_4})
+G^A(\theta,\theta')(\vec{\tau}_{\upsilon_1\upsilon_3}\cdot\vec{\tau}_{\upsilon_2\upsilon_4})(\delta_{\varsigma_1\varsigma_3}\delta_{\varsigma_2\varsigma_4})\nn\\
&&+H(\theta,\theta')(\vec{\tau}_{\upsilon_1\upsilon_3}\cdot\vec{\tau}_{\upsilon_2\upsilon_4})(\vec{s}_{\varsigma_1\varsigma_3}\cdot\vec{s}_{\varsigma_2\varsigma_4}).
\eea
In addition to the usual charge ($F^S$) and spin ($F^A$) components, the Landau function for the multi-valley system contains also valley ($G^A$) and spin-valley components ($H$), which account for inter-valley interactions. In Eq.~\eqref{eq:LFGr}, we neglected the difference between the in- and out-of-plane components of $G^A$ and $H$, which results from processes that swap electrons between different valleys. The matrix element of such processes is small as long as the radius of \emph{eei} is much larger than the lattice spacing, which we assume to be the case here. 

As for the single-valley case, the components of the Landau function can be decomposed into angular harmonics according to $X(\theta,\theta')=\sum_m X_me^{i m(\theta-\theta')}$, where $X\in\{F^S,F^A,G^A,H\}$ and $m$ is angular harmonic. In this case, the expression for current density is: 
\bea\label{eq:jG-int}
J_x&=&e\nu^T_F\left[v_{\rm{F}}f^{S,(1)}\frac{u_0^{(-1)}+u_0^{(1)}}{2}+\frac{\lambda_{\rm R}}{2k_{\rm{F}}}\left( f^{(1)}_+\frac{u^{(-1)}_2-u^{(1)}_2}{2i}-f^{(1)}_-\frac{u^{(-1)}_3+u^{(1)}_3}{2}
\right)\right],\nn\\
J_y&=&e\nu^T_F\left[v_{\rm{F}}f^{S,(1)}\frac{u_0^{(-1)}-u_0^{(1)}}{2i}-\frac{\lambda_{\rm R}}{2k_{\rm{F}}}\left( f^{(1)}_+\frac{u^{(-1)}_2+u^{(1)}_2}{2}+f^{(1)}_-\frac{u^{(-1)}_3-u^{(1)}_3}{2i} \right) \right].
\eea
With Eq.~\eqref{eq:LFGr} taken into account, the equations of motion for electric-field driving are given by
\bea\label{eq:EomGr-intE}
i\Omega u_0^{(m)}&=&-\frac{ev_{\rm{F}}}{2} \Big[(\delta_{m,1} + \delta_{m,-1}) E_x - i (\delta_{m,1} - \delta_{m,-1}) E_y \Big],\nn\\
i\Omega u_1^{(m)}+\lambda_{\rm R}u_2^{(m)}f_+^{(m)}+i\lambda_{\rm R}u_3^{(m)}f^{(m)}_-&=&0,\nn\\
-\lambda_{\rm R} u_1^{(m)}f^{(m)}+i\Omega u_2^{(m)}+\lambda_{\rm Z}M_{3,3}^{(m)}h_+^{(m)}-i\lambda_{\rm Z}M_{3,2}^{(m)}h_-^{(m)}&=&ie\frac{\lambda_{\rm R}}{4k_{\rm{F}}} \Big[(\delta_{m,1} - \delta_{m,-1}) E_x - i (\delta_{m,1} + \delta_{m,-1}) E_y \Big],\nn\\
i\Omega u_3^{(m)}-\lambda_{\rm Z}M_{3,2}^{(m)}h_+^{(m)}-i\lambda_{\rm Z}M_{3,3}^{(m)}h_-^{(m)}&=&0,\nn\\
i\Omega M_{3,1}^{(m)}+\lambda_{\rm R}M_{3,2}^{(m)}h_+^{(m)}+i\lambda_{\rm R}M_{3,3}^{(m)}h_-^{(m)}&=&0,\nn\\
-\lambda_{\rm R}M_{3,1}^{(m)}h^{(m)}+i\Omega M_{3,2}^{(m)}+\lambda_{\rm Z}u_3^{(m)}f_+^{(m)}-i\lambda_{\rm Z}u_2^{(m)}f_-^{(m)}&=&0,\nn\\
i\Omega M_{3,3}^{(m)}-\lambda_{\rm Z}u_2^{(m)}f_+^{(m)}-i\lambda_{\rm Z}u_3^{(m)}f_-^{(m)}&=&0.
 \eea
Same equations for Zeeman-field driving  read 
\bea\label{eq:EomGr-intB}
i\Omega u_0^{(m)}&=&0,\nn\\
i\Omega u_1^{(m)}+\lambda_{\rm R}u_2^{(m)}f_+^{(m)}+i\lambda_{\rm R}u_3^{(m)}f^{(m)}_-&=&\dfrac{g\mu_{\rm{B}}\lambda_{\rm R}}{4} \Big[(\delta_{m,1} + \delta_{m,-1}) B_x - i (\delta_{m,1} - \delta_{m,-1}) B_y \Big],\nn\\
-\lambda_{\rm R} u_1^{(m)}f^{(m)}+i\Omega u_2^{(m)}+\lambda_{\rm Z}M_{3,3}^{(m)}h_+^{(m)}-i\lambda_{\rm Z}M_{3,2}^{(m)}h_-^{(m)}&=&\dfrac{g\mu_{\rm{B}}\lambda_{\rm R}}{2} \delta_{m,0} B_z,\nn\\
i\Omega u_3^{(m)}-\lambda_{\rm Z}M_{3,2}^{(m)}h_+^{(m)}-i\lambda_{\rm Z}M_{3,3}^{(m)}h_-^{(m)}&=&0,\nn\\
i\Omega M_{3,1}^{(m)}+\lambda_{\rm R}M_{3,2}^{(m)}h_+^{(m)}+i\lambda_{\rm R}M_{3,3}^{(m)}h_-^{(m)}&=&0,\nn\\
-\lambda_{\rm R}M_{3,1}^{(m)}h^{(m)}+i\Omega M_{3,2}^{(m)}+\lambda_{\rm Z}u_3^{(m)}f_+^{(m)}-i\lambda_{\rm Z}u_2^{(m)}f_-^{(m)}&=&-i\frac{g\mu_{\rm{B}}\lambda_{\rm Z}}{4} \Big[(\delta_{m,1} - \delta_{m,-1}) B_x - i (\delta_{m,1} + \delta_{m,-1}) B_y \Big],\nn\\
i\Omega M_{3,3}^{(m)}-\lambda_{\rm Z}u_2^{(m)}f_+^{(m)}-i\lambda_{\rm Z}u_3^{(m)}f_-^{(m)}&=&-\frac{g\mu_{\rm{B}}\lambda_{\rm Z}}{4} \Big[(\delta_{m,1} + \delta_{m,-1}) B_x - i (\delta_{m,1} - \delta_{m,-1}) B_y \Big].
 \eea
Here, in addition to Eq. (\ref{eq:fs}) we have defined
\bea\label{eq:hs}
&h^{(m)}=1+H_m,~~~h^{(m)}_{+}=1+\frac{H_{m-1}+H_{m+1}}{2},~~~h^{(m)}_{-}=\frac{H_{m-1}-H_{m+1}}{2}.&
\eea
Like in the single valley case, the renormalization of the couplings (via $f^{(m)}_\pm$, $f^{(m)}$, $h^{(m)}_\pm$ and $h^{(m)}$ factors)  suggests splitting of the resonant frequencies for the in-plane and out-of-plane CSMs. In the two-valley case, these renormalizations also couple the spin-chiral and spin-valley degrees of freedom leading to an additional splitting of resonance frequencies. Solving for the eigenvalues of the systems \eqref{eq:EomGr-intE} and \eqref{eq:EomGr-intB} we get the following resonant frequencies \cite{kumar2021}:
\bea\label{eq:reosnancefreq}
\Omega^2_\pm &=& \Gamma^2\pm\Omega_0^2,\nn\\
\text{where}~\Gamma^2&\equiv&\left(\dfrac{f^{(1)}f_+^{(1)}+h^{(1)}h_+^{(1)}}{2}\right)\lambda_{\rm R}^2+(f_-^{(1)}h_-^{(1)}+f_+^{(1)}h_+^{(1)})\lambda_{\rm Z}^2,\;\text{and}\nn\\
\Omega_0^2&\equiv&\left[\left(\dfrac{f^{(1)}f_+^{(1)}-h^{(1)}h_+^{(1)}}{2}\right)^2\lambda_{\rm R}^4+(f_+^{(1)}h_-^{(1)}+f_-^{(1)}h_+^{(1)})^2\lambda_{\rm Z}^4+(f^{(1)}f_-^{(1)}+h^{(1)}h_-^{(1)})(f_+^{(1)}h_-^{(1)}+f_-^{(1)}h_+^{(1)})\lambda_{\rm R}^2\lambda_{\rm Z}^2\right]^{1/2}.\nn\\
\eea
The response to the electric field driving yields the following current density
\bea\label{eq:jGre}
\bf J&=&\underbrace{\frac{i\sigma_0}{\tau}\left[\frac{f^{S,(1)}}{\Omega }+\frac{1}{\Omega}\left(\frac{\Omega}{2\mu}\right)^2R^{\rm Di}_{\rm JE}(\Omega)\right]}_{\sigma(\Omega)}{\bf E}_0,\nn\\
\text{where}~R^{\rm Di}_{\rm JE}(\Omega)&=&\dfrac{W^{\rm JE}_+}{(\Omega^2-\Omega_+^2)}-\dfrac{W^{\rm JE}_-}{(\Omega^2-\Omega_-^2)},\nn\\
W^{\rm JE}_\pm &=&\dfrac{f_+^{(1)}\lambda_{\rm R}^2}{2\Omega_0^2}[\Omega_\pm^2-\Omega_{\rm JE}^2],~~\Omega_{\rm JE}^2=h^{(1)}h_+^{(1)}\lambda_{\rm R}^2+\frac{[(f_+^{(1)})^2-(f_-^{(1)})^2]h_+}{f_+^{(1)}}\lambda_{\rm Z}^2,
\eea
and magnetization
\bea\label{eq:maggr2}
{\bf M}&=&\underbrace{\dfrac{-i\sigma^{\rm ME}_0}{\Omega\tau}\left(\dfrac{\Omega}{\lambda_{\rm R}}\right)^2R^{\rm Di}_{\rm ME}(\Omega)}_{\sigma^{\rm ME}(\Omega)} {\bf E}_0\times \hat{z};\nn\\
M_z&=&0,\nn\\
\text{where}~R^{\rm Di}_{\rm ME}(\Omega)&=&\dfrac{W^{\rm ME}_+}{(\Omega^2-\Omega_+^2)}-\dfrac{W^{\rm ME}_-}{(\Omega^2-\Omega_-^2)},\nn\\
W^{\rm ME}_\pm&=&\dfrac{\lambda_{\rm R}^2}{2\Omega_0^2}[\Omega_\pm^2-\Omega_{\rm ME}^2],\\
\Omega^2_{\rm ME}&=&h^{(1)} h_+^{(1)}\lambda_{\rm R}^2+(f_+^{(1)}-f_-^{(1)})(h_+^{(1)}-h_-^{(1)})\lambda_{\rm Z}^2.
\eea

\begin{figure*}[htp]
\centering
\includegraphics[width=0.35\linewidth]{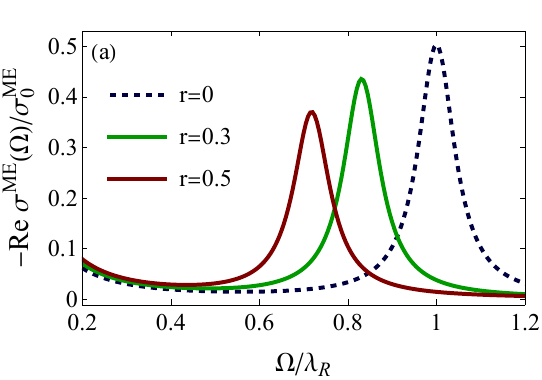}~
\includegraphics[width=0.35\linewidth]{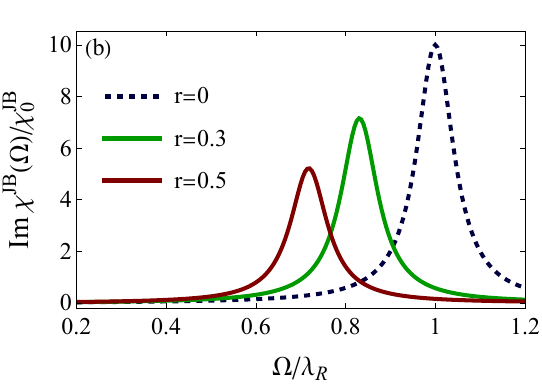}~
\\
\includegraphics[width=0.35\linewidth]{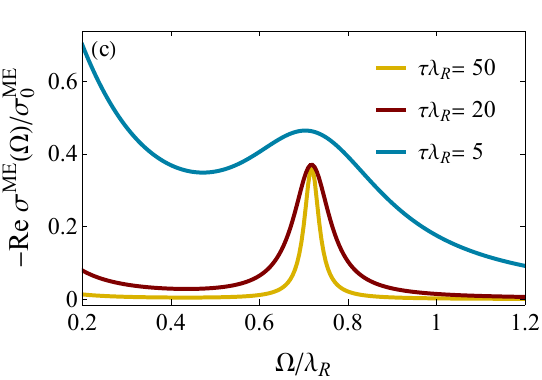}~
\includegraphics[width=0.35\linewidth]{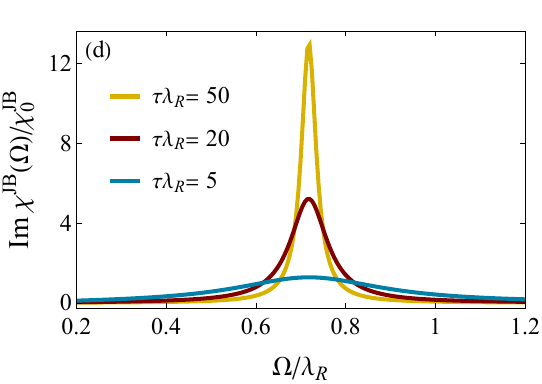}~
\\
\includegraphics[width=0.35\linewidth]{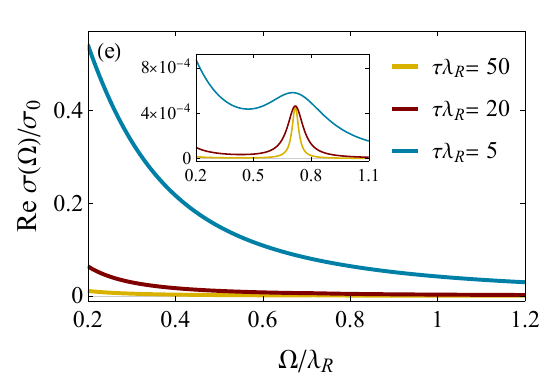}~
\includegraphics[width=0.35\linewidth]{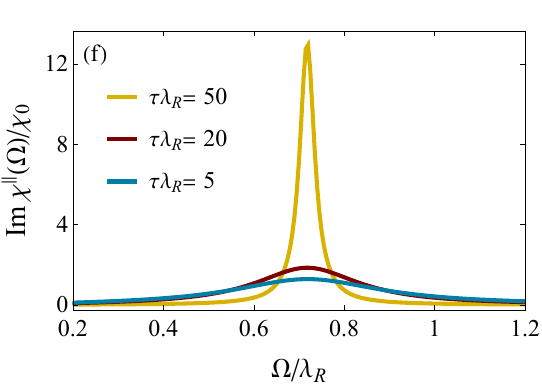}~
\caption{\textbf{Dynamical response of a single-valley system with Rashba spin-orbit coupling.} The cross-response functions,  $\sigma^{\rm ME}$ [panel (a)] and $\chi^{\rm JB}$ [panel(b)] for increasing values of interaction parameter $r$, such that $F^A_0=-r,F^A_1=-r/2,F^A_2=-r/4,F_1^S=r/6$. Here, $\tau\lambda_{\rm R}=20,\mu=5\lambda_{\rm R}$. 
Panels (c) and (d) depict the same response functions as in (a) and (b), respectively, but for different relaxation times, while the interaction parameter is  fixed at $r=0.5$. Panels (e) and (f) depict the direct response functions, the real part of the conductivity (left) and imaginary part of the spin susceptibility (right), for the same set of parameters as in panels (c) and (d). The inset in panel (e) shows the conductivity with the Drude background subtracted.}\label{fig:2DEG}
\end{figure*}

\begin{figure*}[htp]
\centering
\includegraphics[width=0.35\linewidth]{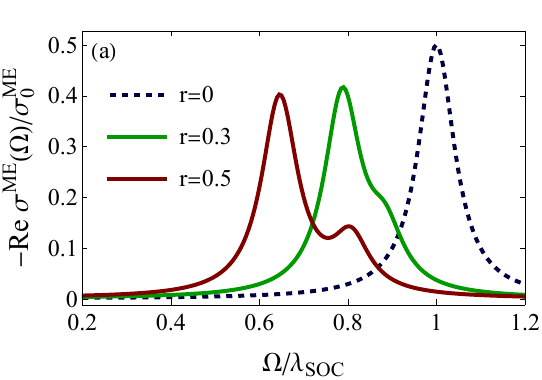}~
\includegraphics[width=0.35\linewidth]{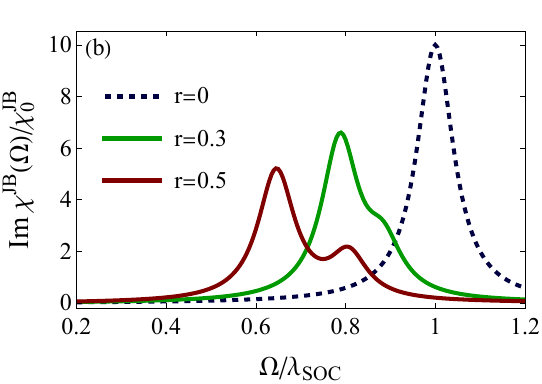}~
\\
\includegraphics[width=0.35\linewidth]{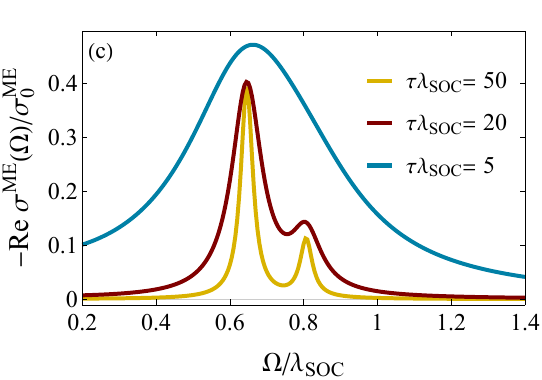}~
\includegraphics[width=0.35\linewidth]{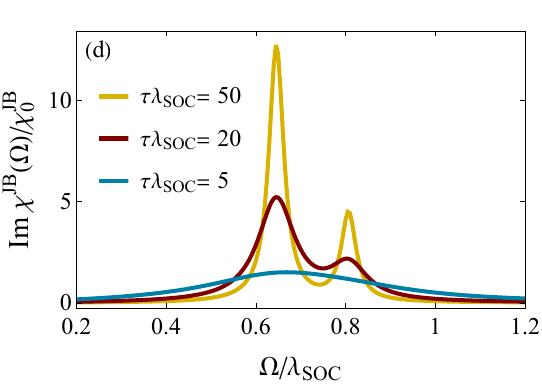}~
\\
\includegraphics[width=0.35\linewidth]{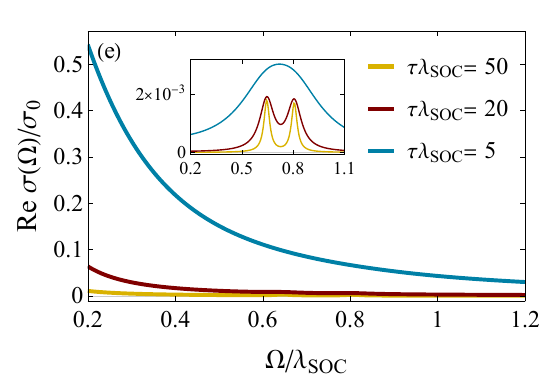}~
\includegraphics[width=0.35\linewidth]{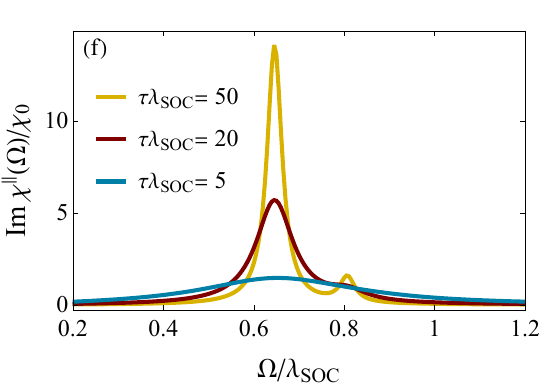}~
\caption{\textbf{Dynamical response of a two-valley system with Rashba and valley-Zeeman spin-orbit couplings.} The cross-response functions, $\sigma^{\rm ME}$ [panel (a)] and $\chi^{\rm JB}$ [panel (b)], for increasing values of interaction parameter $r$, such that $F^A_0=-r,F^A_1=-r/2,F^A_2=-r/4,F_1^S=r/6$, and $H_{0,1,2}=0.9F^A_{0,1,2}$. Here, $\tau\lambda_{\rm SOC}=20,\mu=5\lambda_{\rm SOC}$, $\lambda_{\rm SOC}=\sqrt{\lambda^2_{\rm R}+\lambda^2_{\rm Z}}$, and $\lambda_{\rm Z}=0.5\lambda_{\rm R}$. 
Note that the splitting of the resonance peak increases with interaction. Panels (c) and (d) depict the same response functions as in (a) and (b) but for different relaxation times with fixed interaction parameter $r=0.5$. Panels (e) and (f) depict the direct response functions, the real part of the conductivity (left) and imaginary part of the spin susceptibility (right), for the same set of parameters as in panels (c) and (d). The inset in panel (e) shows the conductivity with the Drude background subtracted.}\label{fig:Di}
\end{figure*}

For Zeeman-field driving the current density is given by
\bea\label{eq:jGrb}
\bf J&=&\underbrace{-\chi_0^{\rm JB}R^{\rm Di}_{\rm JB}(\Omega)}_{\chi^{\rm JB}(\Omega)}{\bf B}_0\times\hat z,\nn\\
\text{where}~R^{\rm Di}_{\rm JB}(\Omega)&=&\dfrac{W^{\rm JB}_+}{\Omega^2-\Omega_+^2}-\dfrac{W^{\rm JB}_-}{\Omega^2-\Omega_-^2},\nn\\
W^{\rm JB}_\pm&=&\dfrac{f^{(1)}f_+^{(1)}\lambda_{\rm R}^2+(f_+^{(1)}+f_-^{(1)})(h_+^{(1)}+h_-^{(1)})\lambda_{\rm Z}^2}{2\Omega_0^2}(\Omega_\pm^2-\Omega^2_{\rm JB}),\\
\Omega_{\rm JB}^2&\equiv&\dfrac{[f_+^{(1)}h^{(1)}\lambda_{\rm R}^2+(f_+^{(1)}-f_-^{(1)})(f_+^{(1)}+f_-^{(1)})\lambda_{\rm Z}^2][f^{(1)}h_+^{(1)}\lambda_{\rm R}^2+(h_+^{(1)}-h_-^{(1)})(h_+^{(1)}+h_-^{(1)})\lambda_{\rm Z}^2]}{f^{(1)}f_+^{(1)}\lambda_{\rm R}^2+(f_+^{(1)}+f_-^{(1)})(h_+^{(1)}+h_-^{(1)})\lambda_{\rm Z}^2},
\eea
while the magnetization reads
\bea\label{eq:MGb}
\bf M&=&\underbrace{-\frac{\chi_0}2R^{\parallel \rm Di}_{\rm MB}(\Omega)}_{\chi^\parallel(\Omega)}{\bf B}_0,\;
M_z=\underbrace{-\chi_0R^{\perp \rm Di}_{\rm MB}(\Omega)}_{\chi^\perp(\Omega)}B_z,\nn\\
\text{where}~R^{\parallel \rm Di}_{\rm MB}(\Omega)&=&\dfrac{W^{\rm MB}_+}{\Omega^2-\Omega_+^2}-\dfrac{W^{\rm MB}_-}{\Omega^2-\Omega_-^2},\nn\\
W^{\rm MB}_{\pm}&=&\dfrac{f^{(1)}\lambda_{\rm R}^2+2(h_-^{(1)}+h_+^{(1)})\lambda_{\rm Z}^2}{2 \Omega_0^2}[\Omega_\pm^2-\Omega^2_{\rm MB}],\\
\Omega_{\rm MB}^2&\equiv&\dfrac{[h^{(1)}\lambda_{\rm R}^2+2(f_+^{(1)}-f_-^{(1)})\lambda_{\rm Z}^2][f^{(1)}h_+^{(1)}\lambda_{\rm R}^2+(h_+^{(1)}-h_-^{(1)})(h_+^{(1)}+h_-^{(1)})\lambda_{\rm Z}^2]}{f^{(1)}\lambda_{\rm R}^2+2(h_+^{(1)}+h_-^{(1)})\lambda_{\rm Z}^2},\nn\\
R^{\perp \rm Di}_{\rm MB}(\Omega)&=&\dfrac{f_+^{(0)}\lambda_{\rm R}^2}{\Omega^2-f^{(0)}f_+^{(0)}\lambda_{\rm R}^2-f_+^{(0)}h_+^{(0)}\lambda_{\rm Z}^2}.\label{eq:MGb2}
\eea
The three resonant frequencies are $\Omega_+, \Omega_-$ and $\Omega_z=\sqrt{f^{(0)}f^{(0)}_+\lambda_{\rm R}^2+f^{(0)}_+h^{(0)}_+\lambda^2_{\rm Z}}$. The first two correspond to the in-plane CSM which is split into two due to the coupling of the spin-chiral and spin-valley degrees of freedom. Both the resonances at $\Omega_\pm$ show up in the cross-responses. The
resonance at $\Omega_z$ corresponds to oscillations of $M_z$ and does not appear in the cross-response. 

\section{Analysis of the results}
\label{sec:analysis}
\begin{figure*}[htp]
\centering
\includegraphics[width=0.85\linewidth]{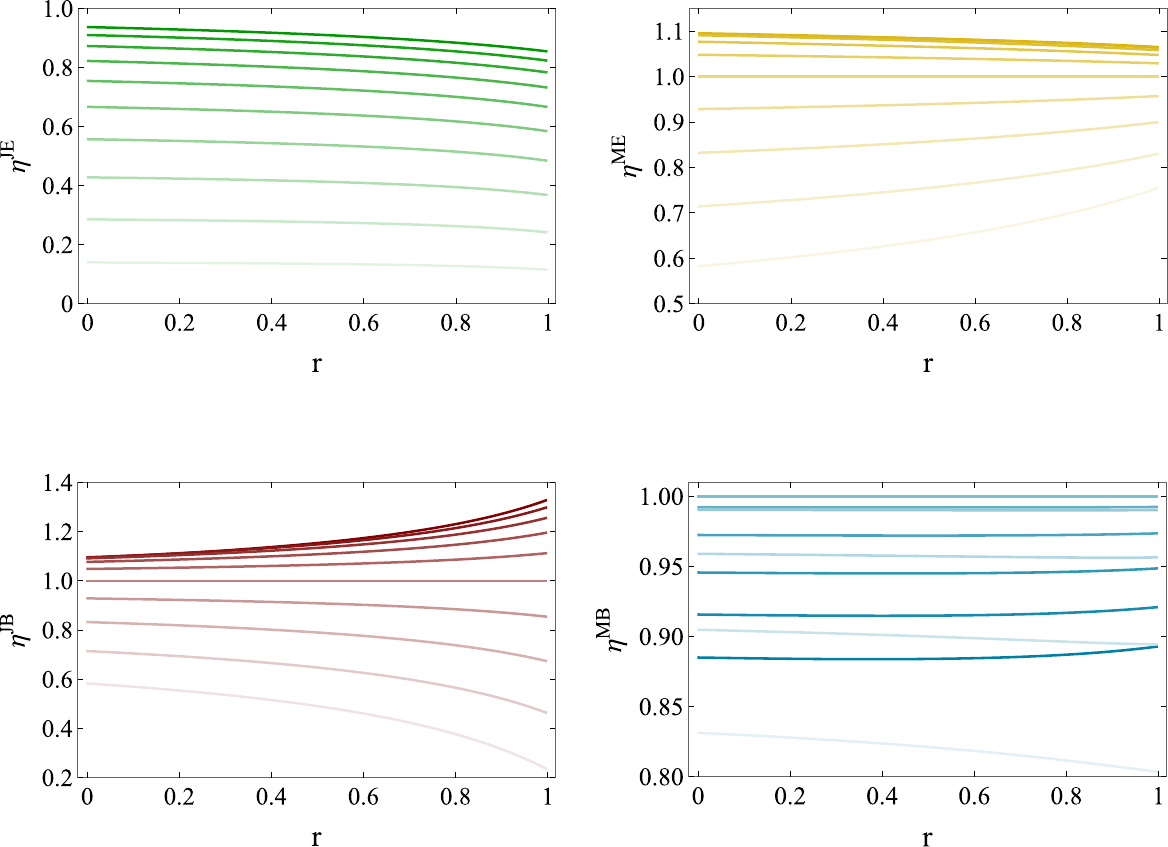}
\caption{\textbf{Ratio of the spectral weights of split resonances in a multi-valley system.} The variation of the ratio $\eta^{AA'}$ with the parameter $r$. The shades from dark to light in each case correspond to $u\equiv H^{(m)}/F^{(m)}\in\{0,0.1,...,0.9\}$. We can see that for the responses ME,JB,MB, the ratio is closer to 1 indicating that the spectral weight is dominated by one mode. For the JE response, this only happens for very weak $u$ and then relatively quickly moves towards $\eta\sim 0$ indicating comparable spectral weights.}\label{fig:Weights}
\end{figure*}

In Figs.~\ref{fig:2DEG}(a), \ref{fig:2DEG}(b) and \ref{fig:Di}(a), \ref{fig:Di}(b) we show the cross-response functions, Re$\sigma^{\rm ME}(\Omega)$ and Im$\chi^{\rm JB}(\Omega)$, derived in the previous two sections, for single- and multi-valley systems, respectively, for a range of interaction parameters. In the single-valley case, we see that the interaction renormalizes the frequencies downwards and also reduces the amplitude of the resonance. For the multi-valley system it does the same but additionally splits the peak into two. This is because of the appearance of a second CSM that involves fluctuations in the spin-valley sector. Observe that while the resonances in $\sigma^{\rm ME}$ and $\sigma$ [Fig.~\ref{fig:2DEG}(c) and inset in Fig.~\ref{fig:2DEG}(e)] become broader as the relaxation time decreases, their heights do not change significantly. This should be contrasted with the behavior in $\chi^{\rm JB}$ and $\chi$ [Figs. ~\ref{fig:2DEG}(d) and (f)] where the resonances are broadened and also suppressed. The usual suppression of the height of the resonance is not seen in panels (c) and (e) simply because $\sigma_0^{\rm ME}$ and $\sigma_0$ are proportional to $\tau$. The same applies to Figs.~\ref{fig:Di} (c)-(f).

Next, in the two-valley case (Fig. \ref{fig:Di}), we observe that the spectral weight is distributed between the two split resonances very unevenly for $\sigma^{\rm ME},\chi^{\rm JB},$ and $\chi$, but evenly for $\sigma$. To elucidate this point, we analyze the ratio of the spectral weights as a function of interaction parameters. According to Eqs.~(\ref{eq:jGre})-(\ref{eq:MGb2}), the spectral weight of a resonance at frequency $\Omega_\pm$ is given by $\pm W^{AA'}_{\pm}/2\Omega_\pm$, where $A=J,M$ and $A'=E,B$. To compare the ratios of spectral weights in various response functions, we introduce a dimensionless 
quantity which is the ratio of the difference in spectral weights of the lower and higher energy modes to their sum:
\bea\eta^{AA'}\equiv \frac{(-W^{AA'}_-/2\Omega_-)-W^{AA'}_+/2\Omega_+}{(-W^{AA'}_-/2\Omega_-)+W^{AA'}_+/2\Omega_+}.\eea Using Eqs. (\ref{eq:jGre})-(\ref{eq:MGb2}), we arrive at
\bea\label{eq:ratioof}
\eta^{AA'}&=&\left(\frac{\Omega_++\Omega_-}{\Omega_+-\Omega_-}\right)\frac{\Omega^2_{AA'}-\Omega_+\Omega_-}{\Omega^2_{AA'}+\Omega_+\Omega_-}.
\eea
We see that across different responses $AA'$, it is basically $\Omega^2_{AA'}/\Omega_+\Omega_-$ [where $\Omega_{AA'}$ are introduced in Eqs.~(\ref{eq:jGre})-(\ref{eq:MGb2})] that controls the spectral weight distribution. The ratio $\eta^{AA'}$ being 
close to one (minus one) would indicate that the spectral weight is predominantly carried by the lower (higher) energy mode, while it being close to zero would indicate comparable spectral weights for both modes. If we set the threshold for this distinction to be at $\sim0.5$, then we see in Fig. \ref{fig:Weights} that ME, JB, and MB responses are dominated by one mode, irrespective of the choices of the interaction, characterized by $r$ (as introduced above) and $u\equiv H^{(m)}/F^{(m)}$. The ratios being positive suggests that the spectral weight of the lower energy mode is higher in the chosen parameter regime in Fig.~\ref{fig:Weights}.  This shows that the lob-sided features of these responses in Fig. \ref{fig:Di} are a generic feature of all responses except JE. In the JE response, there are both regimes. When the inter-valley interaction is very weak ($u<0.3$), the spectral weight distribution in the conductivity is lopsided like in other responses. But as the inter-valley interaction increases, the spectral weights for both modes become comparable. The parameters chosen in Fig. \ref{fig:Di} puts the system in this second regime and hence there are two peaks in $\sigma$ of comparable strength. 

\section{Conclusions}\label{Sec:Conclusion}
In this article, we analyzed the effect of electron-electron interactions on resonances in 
the electric current and magnetization induced by oscillatory electric and magnetic fields in single and multi-valley systems. As an example of the former, we considered a two-dimensional electron gas (2DEG) with Rashba or Dresselhaus spin-orbit coupling (SOC), while as an example of the latter, we considered a monolayer graphene with proximity-induced Rashba and valley-Zeeman SOC. In both cases, a combination of SOC and electron-electron interaction gives rise to chiral-spin modes (CSMs), which are collective excitations of a Fermi liquid with SOC. In addition to direct responses, i.e., optical conductivity $\sigma(\Omega)$ and dynamical spin susceptibility $\chi(\Omega)$ we have shown that 
CSMs also lead to a resonant behavior in the cross-response functions, i.e. magnetoelectric (Edelstein) conductivity $\sigma^{\rm ME}(\Omega)$, which relates the magnetization to an oscillatory electric field, and the current-Zeeman susceptibility $\chi^{\rm JB}(\Omega)$, which relates the electric current to an oscillatory Zeeman field. The last two quantities describe the resonant versions of the Edelstein effect \cite{EDELSTEIN1990} and its inverse \cite{shen2014}. The resonances in cross-responses are induced only by those types of SOCs that couple the vector potential to the electron spin, such as in the Rashba and Dresselhaus types of SOC. Finally, it is shown that out of the three possible CSMs (two with in-plane polarization and one with out-of-plane polarization), it is the in-plane modes that contribute to the cross-responses.

In a single valley system, the effect of electron correlations is to renormalize the resonance frequency. In multi-valley system, correlations also split a single CSM into two, corresponding to coupled excitations in the spin-chiral and spin-valley sectors. While there is a resonant enhancement of the cross-response at both these frequencies, the spectral weight associated with these modes are shown to vary significantly. The higher energy mode is shown to carry a lower spectral weight for a wide range of parameters corresponding to intra- and inter-valley interactions. We add here that although the specific form of the SOC perturbation will depend on the particular system (like the Rashba SOC form for the multi valley system is specific to the Dirac Hamiltonian), the fact that there will be CSMs and that they will be split in the presence of inter-valley interaction remains a generic feature of multi-valley systems, and all of our conclusions should apply.

To conclude, one can state that a general response of spin-orbit coupled metal to oscillatory $E$ and $B$ fields can be written as:
\bea\label{eq:conc}
J_\alpha &=\underbrace{\sigma_{\alpha\beta}(\Omega)}_\text{Drude+EDSR+Hall}E_\beta &+ \underbrace{\chi^{\rm JB}_{\alpha\beta}(\Omega)}_\text{Resonant inverse Edelstein} B_\beta,\nn\\
M_\alpha&=\underbrace{\sigma^{\rm ME}_{\alpha\beta}(\Omega)}_\text{Resonant Edelstein} E_\beta &+~~~~~~~~~~\underbrace{\chi_{\alpha\beta}(\Omega)}_\text{CSR}B_\beta.
\eea
The underbraces below the response functions denote the effects hosted by them. The tensor structure of the cross-response functions depends on the type of SOC [\textit{vide} Eq. (\ref{eq:TensorStructures})] and this fact may be used to identify various types of SOC present in a material. With respect to the polarization of the oscillatory fields, we showed that the in-plane $E$ and $B$ fields induce both direct and cross-responses, while the out-of-plane $B$-field induces only the direct response. These calculations offer a way to track the evolution of the resonances identified in Ref. \cite{MojdehA} and their related effects in different materials. 

\paragraph*{Acknowledgments:} MS and SM were funded by the Natural Sciences and Engineering Research Council of Canada (NSERC) Grant No. RGPIN-2019-05486. A.K. was supported by the FSU Quantum Postdoctoral Fellowship from Florida State University. A.K. also acknowledges support from Canada First Research Excellence Fund and by the Natural Sciences and Engineering Research Council of Canada (NSERC) under Grant No. RGPIN-2019-05312 during his time at Sherbrooke.  DLM was supported by the US National Science Foundation via grant  DMR-2224000. 

\bibliography{References.bib}
\end{document}